\documentclass[sigconf]{acmart}

\AtBeginDocument{%
  \providecommand\BibTeX{{%
    \normalfont B\kern-0.5em{\scshape i\kern-0.25em b}\kern-0.8em\TeX}}}

\copyrightyear{2024}
\acmYear{2024}
\setcopyright{rightsretained}
\acmConference[CSCW Companion '24]{Companion of the 2024 Computer-Supported Cooperative Work and Social Computing}{November 9--13, 2024}{San Jose, Costa Rica}
\acmBooktitle{Companion of the 2024 Computer-Supported Cooperative Work and Social Computing (CSCW Companion '24), November 9--13, 2024, San Jose, Costa Rica}\acmDOI{10.1145/3678884.3681912}
\acmISBN{979-8-4007-1114-5/24/11}

\usepackage{tabularx}
\usepackage{multirow}

\begin{document}

\title[Exploring Role-playing Interactions with Large Language Models to Foster Design Questioning Skills]{Identify Design Problems Through Questioning: Exploring Role-playing Interactions with Large Language Models to Foster Design Questioning Skills}

\author{Hyunseung Lim}
\affiliation{
  \institution{KAIST}
  \country{Daejeon, Republic of Korea}
  }
\email{charlie9807@kaist.ac.kr}

\author{Dasom Choi}
\affiliation{
  \institution{KAIST}
  \country{Daejeon, Republic of Korea}
}
\email{dasomchoi@kaist.ac.kr}

\author{Hwajung Hong}
\affiliation{
  \institution{KAIST}
  \country{Daejeon, Republic of Korea}
}
\email{hwajung@kaist.ac.kr}

\renewcommand{\shortauthors}{Lim, et al.}

\begin{abstract}
  Identifying design problems is a crucial step for creating plausible solutions, but it is challenging for design novices due to their limited knowledge and experience. Questioning is a promising skill that enables students to independently identify design problems without being passive or relying on instructors. This study explores role-playing interactions with Large Language Model (LLM)-powered Conversational Agents (CAs) to foster the questioning skills of novice design students. We proposed an LLM-powered CA prototype and conducted a preliminary study with 16 novice design students engaged in a real-world design class to observe the interactions between students and the LLM-powered CAs. Our findings indicate that while the CAs stimulated questioning and reduced pressure to ask questions, it also inadvertently led to over-reliance on LLM responses. We proposed design considerations and future works for LLM-powered CA to foster questioning skills.
\end{abstract}




\maketitle

\section{introduction}
In the design thinking process, identifying a design problem is a crucial step for deriving plausible solutions within the design's inherent ambiguity and iterative nature~\cite{dorst2001creativity}. However, it requires understanding various user perspectives and balancing creative exploration with precise direction, particularly challenging for novice design practitioners such as students with little knowledge and experience~\cite{dousay2017confessions}. In design education, instructors encourage students to share and argue their problem definitions to help them clarify the problems~\cite{dalsgaard2013design}. Through iterative argumentation with instructors and peers, students learn how to identify design problems by broadening perspectives and addressing logical gaps~\cite{dalsgaard2013design}.

However, during argumentation with instructors who are experts in design, novice design students tend to rely on the instructor's feedback or accept it passively, which may hinder their ability to define design problems independently~\cite{loh2017re, dousay2017confessions}. To address this challenge, previous researchers have proposed an approach that shifts the student's role from responding to feedback to actively engaging in design questioning~\cite{paul2007critical}. Through activities where students ask questions about their design problem statements, students can fill in gaps in their evidence and solidify their design problems as they independently monitor and assess their thinking~\cite{loh2017re}. While questioning is known to be an effective tool for fostering critical thinking~\cite{eris2004effective}, novice design students often face challenges in practicing it independently, frequently finding themselves pondering, ``What should I ask?'' or ``Am I asking the right questions?''

Recent advancements in Large Language Model (LLM)-powered Conversational Agents (CAs) have opened up new educational opportunities, particularly in enhancing argumentation skills by engaging in discourse with these agents~\cite{sabzalieva2023chatgpt}. Various organizations, such as UNESCO~\cite{sabzalieva2023chatgpt}, have suggested the application of LLMs in higher education activities, such as argumentation~\cite{chen2023chatcot} and personal tutoring~\cite{jin2023teach}, by virtue of their ability to generate human-like response. Particularly in creative domains, LLMs have demonstrated their potential to address ill-structured and ambiguous design situations, including offering inspiration, critiquing ideas, and presenting plausible solutions~\cite{sabzalieva2023chatgpt}.

Given the potential of LLM-powered CAs in higher education and design contexts, we aim to explore LLM-powered CAs to support design students in improving their questioning skills for identifying design problems. We developed a research probe for role-playing interaction using ChatGPT's custom instructions~\cite{openai2023gpt4}, allowing students to ask questions from an instructor's perspective to develop design problems. We conducted a preliminary user study with 16 design novice students assigned to identify the design problems in existing radios for a radio redesign project in real-world university design classes, exploring how they interact with an LLM-powered CA to clarify their design problem. Based on our findings, we discuss the design considerations for LLM-powered CAs to support design students in improving their questioning skills.
\section{Methods}
Our preliminary study aims to understand design considerations for LLM-powered CAs that could foster students' questioning skills for identifying design problems. We used a research probe method to investigate these considerations, observing user interactions with the probe to gather insights. Specifically, we designed a role-playing interaction with LLMs, where design students take on the role of instructors and develop their design problem statements through argumentation with the LLMs. This session detailed our methods and the rationale behind our design choice.

\subsection{Role Play-Interaction with LLM for Questioning}
Previous research on design education tools employing CA has aimed at supporting students in acquiring design knowledge rather than enhancing their ability to formulate questions~\cite{peng2023designquizzer}. Conversely, our approach is more centered on fostering students' skills in asking questions, leading us to refer to prior studies in education on questioning. Previous studies~\cite{andreucci2023conceptualizations} have emphasized the importance of creating an educational environment and perspective that actively encourages students to ask critical questions. Educational strategies such as Problem-based Learning (PBL)~\cite{el2019does}, role-play~\cite{ahern2019literature}, and peer review~\cite{el2019does} were suggested to motivate students and instill a critical attitude. Recent research has highlighted new potential for role-play interactions with LLMs due to their ability to mimic specific personas~\cite{junprung2023exploring}. Leveraging this approach, we adopted a role-playing interaction to simulate a question-and-answer session on design problems.

The purpose of our role-playing activity is to enable students, who typically answer questions, to understand and immerse themselves in the perspective of the questioner. Role-playing in education can be categorized into 'Almost Real life,' 'Acting,' and 'Role switch,' with 'Role switch' effectively shifting students to a new perspective~\cite{rao2012exploring}. Instead of having students role-play a peer feedback session, which corresponds to 'Almost Real Life,' we adopted a role-playing interaction where students act as instructors and LLMs as novice designers. This allows students to switch into the role of an instructor and provide feedback from the instructor's perspective, which can be expected to have a similar effect to learning by teaching~\cite{fiorella2013relative}, as it involves taking on the role of an instructor.

\subsection{Design of LLM-powered Research Probe}
We selected ChatGPT~\cite{openai2023gpt4}, developed by OpenAI, and utilized its ~\textit{custom instructions} feature, which enables users to customize ChatGPT for specific purposes. The custom instruction feature aligns with our conversational agent's role-play interaction, as it requires two inputs: ``What would you like ChatGPT to know about you to provide better responses?'' and ``How would you like ChatGPT to respond?'' Referring to the guidelines of OpenAI, we provided instructions for ChatGPT to recognize itself as a novice design student persona (Table \ref{tab:custominstrcution}). We also provided instructions for ChatGPT to identify the user as an instructor in a design course and provide corresponding instructions (Table \ref{tab:custominstrcution}). Next, we designed an initial prompt to simulate our proposed role-play of instructor and student. Referring to the guidelines of OpenAI, we designed the prompt to include the initial design problem identified by the student and the situation in which the instructor provides critique or feedback on that design problem (Table. \ref{tab:initalprompt}).

\begin{table*}
    \centering
    \begin{tabularx}{\linewidth}{X|X}
         \toprule
          \textbf{What would you like ChatGPT to know about you to provide better responses?} & 
          \textbf{How would you like ChatGPT to respond?} \\
        \toprule
        \texttt{I am the professor of your class. The class is in the Department of Industrial Design, where you learn about design thinking. I'll give feedback on your design project. } & \texttt{You are a student who is taking the design class. You are a novice in design. You should not use bullets or lists when answering. Answer only in colloquial language.} \\
        \hline
    \end{tabularx}
    \caption{ChatGPT custom instructions for instructor-student role-play interaction.}
    \label{tab:custominstrcution}
    \vspace{-0.5cm}
\end{table*}

\begin{table*}
    \centering
    \begin{tabularx}{\linewidth}{l}
         \toprule
          \multicolumn{1}{c}{\textbf{Initial Prompts}}\\
        \toprule
        \renewcommand{\arraystretch}{1.2}
        \texttt{You had a project to redesign a radio as a class assignment. You have now identified} \\
        \texttt{the problems of existing radios or the needs of radio users are as follows.} \\ 
        \texttt{Identified Design Problem : \$\{Your Identified Design Problem\}} \\
        \texttt{Please try to answer my (instructor) questions. Do you understand?} \\
        \hline
    \end{tabularx}
    \caption{Initial Prompts for instructor-student role-play interaction.}
    \label{tab:initalprompt}
    \vspace{-0.5cm}
\end{table*}

\subsection{User Study}
\subsubsection{Participants}
We selected an ongoing design class\footnote{To concentrate on the process of defining a design problem by novices with less experience in design studies, we selected a class that was mainly composed of first-year students and required them to participate in a design project.} at our institution and recruited students enrolled in that class to simulate a real-world class-like experience and foster student motivation. We recruited 16 students from that class who were willing to participate. All of our participants had experience using chatGPT, and their demographic information is as follows in Table \ref{tab:demographic}. 

\subsubsection{Procedure}
In our selected design class, students were engaged in a project to redesign a radio, which included the process of defining design problems associated with radios, generating ideas for solutions, and creating prototypes to actualize these solutions. Our study specifically focused on the stage of defining the problems. Here, we asked students to interact with the CA prototype to clarify the problems associated with the radio.

Our study process referred to previous research that employed critical questioning to identify design problems~\cite{loh2017re}. The process began with an orientation session lasting about 10 minutes, during which we explained the objectives and procedures of the study to the students and provided instructions to enable the role-playing with LLMs. We did not give them specific prompts on how to ask questions but encouraged them to think about what questions they would ask in that situation.

The study consisted of three stages, each lasting 20 minutes. Firstly, students wrote a problem statement identifying problems with existing radios. Second, they conducted a role-play interaction with the CA and asked questions about the design problems they had previously identified. The interaction was initiated by students entering the instructions we provided. Then, they begin the role-play by asking the CA a question. After the role-play interaction, students wrote a new problem statement to re-clarify the design problem based on this activity. After completing all stages, students were interviewed briefly to reflect on their interactions with the prototype. We collected the interaction logs with the LLM and the interview records for analysis.

\subsection{Analysis}
Firstly, for the role-playing interaction data, we classified the user inputs collected according to the design question taxonomy suggested by Eris~\cite{eris2004effective}. This taxonomy categorizes design questions to understand their nature, which are categorized into Low-level Questions (LLQs), Deep Reasoning Questions (DRQs), and Generative Design Questions (GDQs). Previous research has shown that DRQs and GDQs are likely to be more useful questions than LLQs in the process of defining problems and generating new ideas~\cite{eris2004effective, marbouti2019written, hurst2019comparing}, and since design needs iterative processes of divergence and convergence, both DRQs and GDQs can be essential questions in the process of defining problems~\cite{eris2004effective, marbouti2019written}. The first and second authors independently labeled the user inputs according to Eris's taxonomy, then compared their labels, discussed any disagreements, and reached a final categorization.

Secondly, the interview data were qualitatively analyzed using thematic analysis~\cite{virginia2006thematic}. Our analysis focused on the opportunities and challenges of LLM-powered CAs. The first author of this study repeatedly reviewed the open-coded interview transcripts and interaction logs using Atlas.ti. Subsequently, the entire research team discussed and identified patterns and themes during multiple group meetings.

\begin{table}
    \renewcommand{\arraystretch}{1.4}
    {\footnotesize
    \begin{tabular}{ccccc}
        \toprule
        \textbf{\begin{tabular}[c]{@{}c@{}}Participant \\ ID\end{tabular}} & \textbf{Age} & \textbf{Gender} & \textbf{\begin{tabular}[c]{@{}c@{}}Duration of \\ Learning Design\end{tabular}} & \textbf{\begin{tabular}[c]{@{}c@{}}Language \\ used\end{tabular}} \\ \toprule
        P1 & 24 & M & 3years & KR\\ \hline
        P2 & 19 & M & less than a year & KR\\ \hline
        P3 & 18 & F & less than a year & KR\\ \hline
        P4 & 19 & M & less than a year & KR\\ \hline
        P5 & 19 & F & less than a year & KR\\ \hline
        P6 & 19 & M & less than a year & KR\\ \hline
        P7 & 18 & M & less than a year & KR\\ \hline
        P8 & 19 & M & less than a year & EN\\ \hline
        P9 & 18 & M & less than a year & EN\\ \hline
        P10 & 22 & F & less than a year & EN\\ \hline
        P11 & 19 & F & less than a year & EN\\ \hline
        P12 & 19 & F & 2years & EN\\ \hline
        P13 & 21 & M & less than a year & EN\\ \hline
        P14 & 21 & F & a year & EN\\ \hline
        P15 & 21 & F & 2years & EN\\ \hline
        P16 & 19 & F & less than a year & EN\\ \bottomrule
    \end{tabular}
    }
    \caption{Demographic information of students.}
    \label{tab:demographic}
    \vspace{-0.5cm}
\end{table}
\section{findings}
All participants in our study interacted with a customized LLM and generated their own questions. They wrote a total of 172 inputs, averaging 10.8 inputs (SD = 4.5), with each entry consisting of approximately 13.6 (SD = 10.4) words. Of the 156 questions, excluding the 16 non-question inputs, 53 (SD=2.4) were LLQs, 43 (SD=2.3) were DRQs, and 60 (SD=2.3) were GDQs. Even though our participants were novices, they asked many questions that were not LLQs, but GDQs outnumbered DRQs.

\subsection{Benefits of Leveraging LLMs to Foster Questioning from Students}

\subsubsection{Breaking barriers in asking critical questions}
Most students (13/16) appreciated the argumentation with the LLM as it allowed them to ask questions without the fear of being judged. As noted in previous studies~\cite{abdullah2012student}, even though students were encouraged in classes, they were hesitant to ask questions because they felt pressure to ask insightful questions conscious of teacher and peer evaluation. They noted that the non-judgmental nature of our proposed LLM interaction allowed them to start asking questions, even when they lacked confidence. Furthermore, students (4/16) noted the advantage of the LLM not being a real person, which allowed them to ask more critical questions without fearing their questions being perceived as aggressive. P1 said, \textit{``Since LLM is not a real person, I tried to ask questions as critically as possible without feeling guilty and with the intention of making it cry.''} This aspect diminished the concerns about interpersonal dynamics, enabling a more critical and uninhibited approach to questioning.

\subsubsection{Developing Questions via LLM Answers as a Quality Indicator}
Many students (10/16) initiated the role-play interaction with LLM by asking low-level questions about the radio. For example, they asked questions such as \texttt{``When will the radio be used? (P2)''} and \texttt{``What are the advantages of radio? (P5)''} However, they realized that more critical questions were needed to obtain insightful responses in clarifying design problems. In this process, students began to think about how to ask critical questions that could stimulate thinking, just like real instructors.

Students attempted to develop more specific questions on their own through an iterative process of evaluating LLM's responses and determining whether their questions were appropriate. As one approach to asking questions, some students (6/16) reflected on previous experiences in which they had received feedback from their instructor. For example, P15 said, \textit{``At first, I found it challenging to ask questions, often asking broad ones, which led to the GPT providing long-winded and predictable responses. Then, I suddenly remembered the questions my professor was asking. My professor would first commend the strong aspects of my ideas and then critique the weaker parts. Inspired by this, I tried to ask in a similar style.''} Other students also mentioned that they were able to think about how to ask questions, what questions are needed in this context, and what questions to ask to lead students in a good direction. Due to these lessons from students, students initially asked LLQ, but the proportion of high-order questions gradually increased. This process prompted students to reflect on the nature and standard of questioning, drawing from previous feedback sessions.

\subsection{Challenges Students Faced in Role-Playing}
\subsubsection{Repeating Single Pattern Questions}
We noticed that some students (4/16) kept asking the same type of questions over and over again. For example, P7 and P8 kept asking only GDQs, and P9 kept asking only DRQs. The dialogues of students who only asked GDQs kept leading to discussions about finding another problem or finding a solution to the problem, even though the task was to define the problem with the radio. In contrast, P9's dialog was not focused on finding a design problem but rather on writing a good description of the initially defined problem. As a result, instead of asking questions, he would ask, \textit{``This answer seems like you have simply arranged the former answer. I don't like it. Try to focus on the common grounds of the purposes and find the key user needs''.} These students mentioned that they did not know how to ask the question, so they stuck to their guns.

\subsubsection{Over-reliance on LLMs}
Despite LLM's role as novice design students, LLM provided well-argued responses without hesitation due to its extensive design knowledge. This made it challenging for students to come up with critical follow-up questions. Therefore, students gave up arguing with the LLMs through questioning and instead requested them to write a complete problem statement or propose a design idea: \texttt{``How should the appearance of the radio change to solve the problems?''} or \texttt{``Blend these two needs to redefine the problem for radios.''} Among the 16 students, 14 generated these questions during the role-playing interactions. They became reliant on the LLM, which consistently generated plausible responses, leading them to engage less in critical thinking on their own.

\subsubsection{Difficulties in associating argumentation with design problems}
While the argumentation with LLM trained students to pose critical questions, they encountered challenges in clarifying the design problem based on these dialogues. Students recognized the needs of potential users or feasibility issues by asking questions, but many of them (10/16) struggled to integrate these discussions into an organized problem statement. We found that one reason for this is that design problems need to leave space for creative solutions, but students focus on criticizing without reflecting on the context of the design. In particular, students found it difficult to explicitly know which questions were key to defining the problem because LLM was receptive to their questions. In this regard, students noted that the proposed interaction, heavily focused on logical question-and-answer, was difficult to utilize to clarify problems that could lead to deriving design solutions.


\section{Discussion and Design Implications}
\subsection{Adjusting LLM's Knowledge Boundaries to Fit Novice Design Students Role}
We found that the answers generated by LLMs based on their extensive pre-trained knowledge hinder students from perceiving LLMs as fellow students and impede their training in critical questioning. To reduce the unsuitable capabilities of LLMs in their designated roles, it is necessary to control the knowledge statement of LLMs and to ensure that argumentation develops only through students' questions. Previous studies utilizing LLMs as debate partners have emphasized the need to control knowledge statements, proposing prompt engineering and techniques~\cite{jin2023teach}. Particularly in the design context, there is a need for future work on what knowledge should be limited when LLMs perform the role of design students and how this should be implemented.

\subsection{Encouraging Reflection on Questions in Real-time and Guiding to Suitable Questions}
Through the role-play interaction with LLMs, students had the opportunity to think about how to ask critical questions by engaging themselves in the role of an instructor. However, because students were unfamiliar with asking questions, they had no choice but to rely on their previous feedback experiences. For students to improve their questioning skills, it is necessary not only to provide an environment where students can ask questions but also to provide educational support on what critical questions students need and how to ask essential questions at the relevant design thinking stage. In our findings, since students kept repeating the same type of questions because they could not reflect on what type of questions they were asking, it is necessary to enable them to reflect. Recent studies have explored interactions that simulate conversations and allow students to reflect on their own conversations~\cite{shaikh2024rehearsal, jin2024teach}. One direction for future research could be to explore ways to help students reflect on their own questions and ask appropriate questions by analyzing the current question and showing the results. In particular, it can be used to recognize whether a student continues to ask only divergent questions (GDQ) or only convergent questions (DRQ).

\subsection{Engaging Students Ask Questions that Reflect Nature of Design}
In our user study, students posed various critical questions through interactions with LLMs. However, these queries were not adequately reflected in their design problems. Prior studies have highlighted that design is a co-evolving process involving both problem identification and idea generation. Therefore, it is crucial for students to consider potential solutions while posing critical questions consistently. Effective design problem identification should involve discussion that addresses crucial design considerations, such as the problem's relevance to actual users and its solvability. It may be necessary to guide LLMs using existing design criteria or heuristic evaluation metrics and prompt them to ask relevant questions. Consequently, the development of LLM-powered CAs that support design problem identification should focus on enabling students to integrate design considerations aligning with the intrinsic dynamics of design.

\section{conclusion and future work}
In our study, we explore design considerations of LLM-powered conversational agents to support design students to ask questions to identify design problems independently. Our future work is to design and validate an LLM-powered CA that allows critical questions from students based on design implications identified through our research. In this work in progress, we have yet to assess how well an LLM-powered CA elicits critical questions from students compared to traditional methods and its long-term educational effectiveness. Future research will involve designing and testing the performance of LLM-powered CAs and discussing their potential application in design education settings.

\begin{acks}
We thank our participants for their engagement and the anonymous reviewers for their thoughtful comments and suggestions. We also thank professor Min-Kyu Choi for allowing us to use his class time to conduct the experiment.
\end{acks}

\bibliographystyle{ACM-Reference-Format}
\bibliography{contents/10biblography}
\end{document}